\documentclass[12pt]{article}

\usepackage{anyfontsize}         
 
\usepackage{textcomp,marvosym}
\usepackage[english]{babel}
\usepackage[utf8]{inputenc}
\usepackage{amsmath}
\usepackage{amsfonts}
\usepackage{graphicx}
\usepackage[colorinlistoftodos]{todonotes}
\usepackage{dot2texi}
\usepackage[ruled,vlined]{algorithm2e}
\usepackage{enumitem}
\usepackage{fixltx2e}
\usepackage{algpseudocode}
\usepackage{mathtools}  
\usepackage{amsmath}
\usepackage{amssymb}
\usepackage{tabulary}
\usepackage{booktabs}
\usepackage{rotating}
\usepackage{colortbl} 
\usepackage{xcolor} 
\usepackage{xfrac}
\usepackage{subcaption}
\usepackage{array}
\usepackage{xcolor} 
\usepackage{dcolumn}
\usepackage{tabu}
\usepackage{array}
\usepackage{booktabs}
\usepackage{multirow}
\usepackage{tikz}
\usetikzlibrary{arrows.meta,positioning}
\newtheorem{definition}{Definition}[section]
\usetikzlibrary{calc}
\usetikzlibrary{snakes}
\usepackage{cite} 
\usepackage{microtype}

\usepackage{soul}

\DisableLigatures[f]{encoding = *, family = * } 

\newcommand\defeq{\stackrel{\smash{\scriptscriptstyle\mathrm{def}}}{=}}

\newlength\savedwidth

\makeatletter
\newcommand{\doublewidetilde}[1]{{%
  \mathpalette\double@widetilde{#1}%
}}
\newcommand{\double@widetilde}[2]{%
  \sbox\z@{$\m@th#1\widetilde{#2}$}%
  \ht\z@=.9\ht\z@
  \widetilde{\box\z@}%
}
\makeatother

\MakeRobust{\Call}

\def\ErdosRenyi{{Erd\H{o}s-R\'{e}nyi }}

\newcommand{\head}[1]{\textnormal{\textbf{#1}}}
\newcommand{\normal}[1]{\multicolumn{1}{l}{#1}}

\newcolumntype{A}{D{.}{.}{2.3}}

\usepackage[numbers]{natbib}

\newif\ifnotesw \noteswtrue

\usepackage{amsmath} 
\usepackage{amsfonts} 
\usepackage{amssymb}
\usepackage{booktabs}
\usepackage{dot2texi}
\usepackage[ruled,vlined]{algorithm2e}
\usepackage{mathtools}
\usepackage{hyperref}

\begin{document}

\title{
Epidemics on Networks: Reducing Disease Transmission Using  Health Emergency Declarations  and Peer Communication \\
 }
\author{ Asma Azizi$^{1,2, \thanks{Corresponding author.  E-mail address: aazizibo@asu.edu}}$ , Cesar Montalvo$^{1, 2,\dagger }$, Baltazar Espinoza$^{1, 2 }$,  Yun Kang${^ {1,3,\ddagger} }$, \\ and Carlos Castillo-Chavez$^{1, 2, \S }$
 \\
} 

\maketitle 

\begin{center}
\vskip -0.05in
$~^{1}$ School of Human Evolution and Social Change; Simon A. Levin Mathematical Computational Modeling Science Center, Arizona State University, Tempe, AZ 85281\\

$~^{2}$ Division of Applied Mathematics, Brown University, Providence RI,  02906\\
$~^{3}$ Sciences and Mathematics Faculty, College of Integrative Sciences and Arts, Arizona State University, Mesa, AZ 85212, USA\\
\end{center}

\begin{abstract}

Understanding individual decisions in a world where communications and information move instantly via cell phones and the internet, contributes to the development and implementation of policies aimed at stopping or ameliorating the spread of diseases. In this manuscript, the role of official social network perturbations generated by public health officials to slow down or stop a disease outbreak are studied over distinct  classes of static social networks. The dynamics are stochastic in nature with individuals (nodes) being assigned fixed levels of education or wealth. Nodes may change their epidemiological status from susceptible, to infected and to recovered. Most importantly, it is assumed that when the prevalence reaches a pre-determined threshold level, $P^*$, information,  called awareness in our framework, starts to spread, a process triggered by public health authorities. Information is assumed to spread over the same static network and whether or not one becomes a {\it temporary} informer, is a function of his/her level of education or wealth and epidemiological status. Stochastic simulations show that threshold selection $P^*$ and the value of the average basic reproduction number impact the final epidemic size differentially. For the \ErdosRenyi and Small-world networks, an optimal choice for $P^*$ that minimize the final epidemic size can be identified under some conditions while  for Scale-free networks this is not case.

\textit{Keywords:  Awareness spread, Behavior change, Outbreak and epidemic threats, \ErdosRenyi  network, Small-world Network, Scale-free Network,}   
\end{abstract}

\newpage \setcounter{tocdepth}{4}


\section{Introduction} \label{Introduction}
The decisions that individuals make over an epidemic outbreak depend on multiple factors. Here, they are assumed to depend on  available information,  misinformation,  and  the  income/education of those  making  them \cite{fenichel2011adaptive, herrera2011multiple,del2005effects,towers2015mass,perrings2014merging}.  There are multiple possible scenarios that consider the decisions that individuals may  make over the course of an outbreak.  Individuals may modify their behaviors in order to reduce their environmental  susceptibility to a disease by washing their hands frequently, avoiding handshakes and avoiding kissing salutes, not taking public transportation during rush hours, using masks and more.  The frequency and effectiveness of these decisions may depend on the perceived risk of infection, a function of what each individual ``knows". In short, individual responses to new circumstances are adaptive  and may depend  on  real or perceive risks of infection. Current   disease prevalence, may become a  marker or a tipping point, that when crossed, triggers individual or {\it policy} decisions. Whether or not individuals follow public health officials' recommendations may be a function of individuals' economic/educational status.  The determination to make a drastic decision may be weakened or  reinforced by each individual's networks of friends.  Responses are altered by the  opinions of work-related connections. Personal needs  play a role, and they include the need to use public transportation or  the desire to attend a social event.\\

The landscape where behavioral decisions take place is  not fixed.  On the contrary, it may be altered from within (individual decisions) or by the use of preventive or active public health policy decisions or recommendations,  some ``obvious" like vaccination or quarantine, others  drastic like mandated social distancing, as in the 2009-10 flu pandemic in Mexico \cite{herrera2009multiple}.  Now, whether the dynamical  changes experienced by the socio-epidemiological landscape are slow or fast, will depend on many factors.  It is  within this,  often altered, complex adaptive dynamical  system, that individual decisions such as individually-driven social distancing,  the use of face masks, frequent hand washing,  increased condom use, and or the routine use of  non-pharmaceutical interventions (NPIs), takes place \cite{wang2015coupled}. It has been theoretically documented \cite{funk2009spread,perra2011towards, funk2013talk,hyman2007infection,misra2011modeling, tracht2010mathematical}  that massive behavioral changes can impact the patterns of infection spread, possibly playing a  critical role in efforts to prevent or ameliorate  disease transmission. Today policies  are implemented regardless of our knowledge of who are the ``drivers"  responsible for  inducing behavioral changes.   Modeling frameworks exist that allow for the systematic exploration of possible scenarios. Through a systematic exploratory analyses of appropriately selected classes of scenarios, it is possible to identify possibly  effective (model-evaluated) public health policies. The use of highly detailed models including individual-based models has some advantages since they can incorporate  individuals' awareness of risk based on available of  local information. The evaluations carried out on the effectiveness of  changes, at the individual level, can be  used  to assess, for example, the impact of NPIs in reducing disease prevalence. \\

Models that couple  disease dynamics and awareness to  levels of infection risk have been proposed. These models have been used to explore the impact of behavioral changes on the  spread of infection.   In the review paper of Wang et al.  \cite{wang2015coupled}  classify  models as Rule-Based Models (RBMs),  those where individuals make their decision about changing behavior independently of others, and  Economic-Epidemiology models (EE models), that is, models  where individuals  change their behavior in order to maximize their own utility function (what they value) subject to  available  resources. The EE models  account for the responses that individuals  take in response to infection risks on disease prevalence at the population level.   The modeling and results reported in this manuscript are more closely related to those used in RBMs.   \\

Ruled-Based Models \cite{hyman2007infection, epstein2008coupled,sahneh2012existence, perra2011towards,kiss2010impact,misra2011modeling,poletti2011effect,poletti2012risk,poletti2009spontaneous} vary from compartmental ODE models \cite{hyman2007infection,misra2011modeling, tracht2010mathematical} to individual-based network models \cite{wu2012impact,funk2009spread,granell2013dynamical,meloni2011modeling}. These models have been used to study the dynamics of highly diverse diseases including, for example,  influenza and HIV \cite{poletti2011effect, fraser2004factors}, or  in the study of  generic infections \cite{perra2011towards, kiss2010impact, misra2011modeling, poletti2009spontaneous, poletti2012risk}.  Compartmental models  (often using a phenomenological approach)  categories  designed to capture levels of awareness of infection.  Such approach that can be used to incorporate `awareness' in  network models, is the  objective of this manuscript.  Some models  assume that ``awareness''  spreads along with the invading disease, that is, through  identical contact networks. Here, it is assumed that the disease and information spread over the same social network (a drastic simplification).  The possibility that   awareness and responses  to the presence of a new infection  among a subset of the population at risk, may significantly alter  regular   temporal patterns of disease prevalence (lower highs) have been studied.   Studies have  also shown that epidemic thresholds can be altered \cite{sahneh2012existence, poletti2009spontaneous} \cite{perra2011towards} in response to the effectiveness of NPI. Here, the focus is on the role of policy decisions/recommendations in altering disease dynamics, possibly the final epidemic size, within a model where awareness (generated by official actions) spreads among those susceptible to infection and their `friends'. \\

In this paper, we explore the impact of various, prevalence-dependent pre-selected thresholds (decisions made by public health officials)  as triggers of possibly temporarily behavioral change.   The time  to a triggering event is assumed to depend on the prevalence of infection-- a decision taken by health authorities (declaring some level of health emergency). We carry out simulations to explore the impact of variations on triggering prevalence-driven levels on the final epidemic size  of a  non-fatal infection under three distinct  fixed artificial social structures modeled as,  Erdos-Raynei,  Small-world,  and Scale-free networks. 

\section{Modeling Framework}\label{method}

The model captures changing disease  transmission dynamics by incorporating, after a health emergency has been declared.  The spread of  risk information, that play by assumption over the same social network. Risk information is transmitted at different rates depending on the  economic/education levels of individuals.  We  focus on  single outbreaks,  within a Susceptible–Infected–Recovered (SIR) framework. The  dissemination of awareness (risk information) among the susceptible population, a process triggered by the declaration of a health emergency  by the public health authorities, generates temporary  changes on the knowledge and understanding of risk  of infection, thus altering their activity and effectiveness as communicators of risk. Becoming aware of the risk of infection and inciting our ability to communicate risk,  takes place at a rate modeled as a function of the economic/educational level of  individuals in the network. Economic/educational levels are preassigned qualities  to  individuals in the network from  pre-selected distributions.  A susceptible individual may be  a member of three  sub-classes of susceptible: \textbf{U}naware,  $S^u$, \textbf{A}ware, $S^a$ and \textbf{I}ndifferent, $S^i$.  Unaware  individuals  that become aware of the risk of infection  may (depending on their economic/educational level), change their  state to \textbf{A}ware, that is, be ready to  convey information, with different levels of enthusiasm, on the risks of the infection to neighbors in the network. \textbf{A}ware, is assumed to be a temporary state and so a transition to the \textbf{I}ndifferent state is included. The triggers that determine or drive awareness state are a function of the disease prevalence level and not the time since start of the outbreak. The case when delays in transmitting  real-time prevalence information to public health officials takes place, the most likely scenario  is not considered. However, we know that it  brings additional consequences (see \cite{velasco1996effects}). \\

\begin{figure}[h]
    \centering
    \includegraphics[scale=.5]{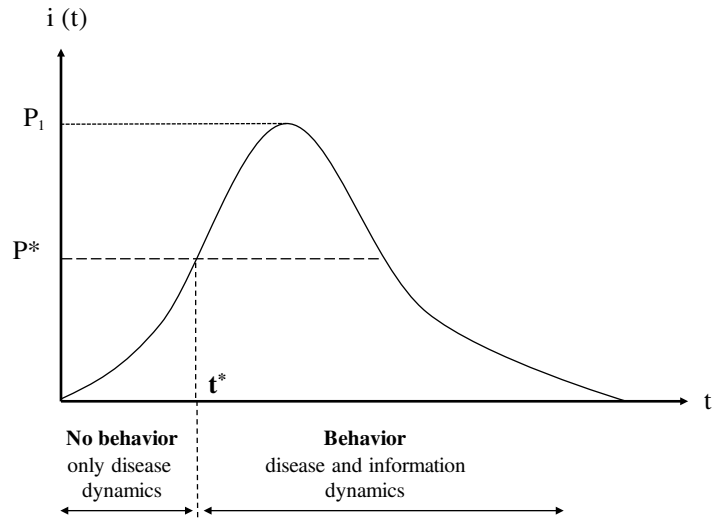}
    \caption{Flow diagram of the model.}
    \label{fig:prevalencetime}
\end{figure}

The activation process of public awareness takes place within a time framework determined by the actual (reported in general) disease prevalence and so, it is a function of the epidemic outbreak growth rate.   We let $P^*$ denote the pre-selected level   of prevalence identified by  public health officials as the triggering threshold after which, a public awareness process campaign is initiated via mass media, word of mouth, social media and the like. The timing is selected when, according to public health officials.  A disease outbreak may be about to reach unacceptable levels, with  $P^*$, being  the pre-selected tolerance level before a health emergency   is declared.  The  $P^*$ determines the time $t^*$ when the information campaign  starts  (see  Figure \ref{fig:prevalencetime}).   The $t^*=t^*(P^*)$ identifies the  public health system threshold that forces public health  action aimed at  reducing outbreak consequences. The actual changes in behavior are determined by an stochastic process, and so, the shift in behaviors, is a function of ``chance''.  \\

Before $t^*$,  the dynamics are governed only by the  disease transmission  process. At the time $t^*$,   the model activates information process that models to  behavioral change that reduce disease transmission. Therefore,  $t^*$ denotes the time when prevalence reaches  $P^*$ for the first time, Figure(\ref{fig:prevalencetime}).\\

 Sub-section \ref{network}  introduces the details of our network model, starting with the network structure and node attributes that are used to represent individuals in the network. 
 
\subsection{Network Structure}\label{network}
The network is denoted by $\textbf{G}=(\textbf{V},\textbf{E})$ which includes a set  $N = |\textbf{V}|$  of nodes representing  individuals $\textbf{V} = \{\textbf{i}| \textbf{i} = 1, 2, 3,  \dots N\}$ together with a binary adjacency relation defined by the set of edges $\textbf{E} = \{\textbf{ij}|  \textbf{i,j} =1,2,3, \dots N\}$, where $\textbf{ij}$ denotes the edge between individual (node) $\textbf{i}$ and individual (node) $\textbf{j}$. 
We  make use of three network structures, namely, \ErdosRenyi, Small-world, and Scale-free networks -defined in the Appendix- as models of our social landscape, the place where infection and  awareness spread and behavioral change  take place.\\

After generating the network \cite{erdds1959random, newman1999renormalization, bollobas2003directed}, we assign some attributes to the nodes (individuals) as follows:  each node 
\textbf{i} is associated with a random variable $x_i\in\left[0,1\right]$, generated from  a beta-distribution, denoting the level of education of individual \textbf{i}, values closer to $0$ corresponding to higher levels of education while those close to $1$  indicate limited education,   an individual \textbf{i} is assigned a random number  from a beta distribution with shape parameters  $\alpha$ and $\beta$. \\

The underlying network \textbf{G} is weighted, where the weight $0<c_{ij}\leq 1$ for edge $\textbf{ij}$ is the probability, of physical contact between neighbors \textbf{i} and \textbf{j} on any specific day.  We assume that  the values $c_{ij}$ are randomly  generated from a uniform distribution $\mathcal{U}[0.5,1]$. For example  if  $c_\textbf{ij}=\frac{5}{7}$, that means that  these two neighbors  meet with probability $\frac{5}{7}$ and do not meet with probability $\frac{2}{7}$ on a specific day. We also assume that there are no birth or deaths in the population of nodes or edges, during the epidemic outbreak. \\

\subsection{Dynamic of awareness among susceptible individuals}
At any given time $t$, a person \textbf{k} is classified as  susceptible, $S_\textbf{k}(t)$,   infected, $I_\textbf{k}(t)$, or  recovered, $R_\textbf{k}(t)$, to the disease under consideration.
 Disease risk  awareness   among susceptible  individuals spreads after public health officials make the deliberate decision to stress vigorously the risk of infection, as it may be the case under an influenza epidemic or other  health emergency. The model assumes that the decision to promote risk prevention starts once a pre-determined prevalence level $P^*$, at time $t^*$,  has been reached. 
Health authorities start to spread   information about a pathogen's outbreak, information that gets communicated by those individuals, actively aware, in the network, with different degrees of efficiency. The spread of information is a function of chance (probability model) and economical/educational level. The state of active information-communicator is assumed to be temporary, with individuals  moving into a``passive" state,  {\it indifferent}, after a short period of time.  Before time $t^*$ every susceptible individual is unaware of the information campaigns but  for times $s>t^*$,    members of $S_\textbf{{k}}(s)$ may change  status from unaware ($S_\textbf{k}^u(s)$) to   aware ($S_\textbf{k}^a(s)$),  then moving, with some probability,  into the indifferent class  ($S_\textbf{k}^i(s)$).\\

We define $\widetilde{\lambda_{jk}}$ as the  probability that $S_\textbf{k}^a(s)$ informs $S_\textbf{j}^u(s)$ at time $s$, and causes  $S_\textbf{j}^u(s) \overset{\widetilde{\lambda_{jk}}}{\rightarrow} S_\textbf{j}^a(s+1)$. Thorough the probability $\widetilde{\gamma_k}$ -generated from an exponential distribution with parameter $\widetilde{\gamma}$, the average awareness period  is therefore $1/\widetilde{\gamma}$ days.  The members of $S_\textbf{k}^a(s)$  leave the awareness class into indifferent class, that is,  $S_\textbf{k}^a(s) \overset{\widetilde{\gamma_k}}{\rightarrow} S_\textbf{k}^i(s+1)$, with probability $\widetilde\gamma_k$,  after spending, on the average  $1/\widetilde{\gamma}$ days  in the aware class; the mean of   waiting time distribution   for the class $S_\textbf{k}^a(s)$.

\subsubsection{Awareness transmission probability}
\begin{figure}[htp]
    \centering
    \includegraphics[width=.65\textwidth,height=.35\textheight]{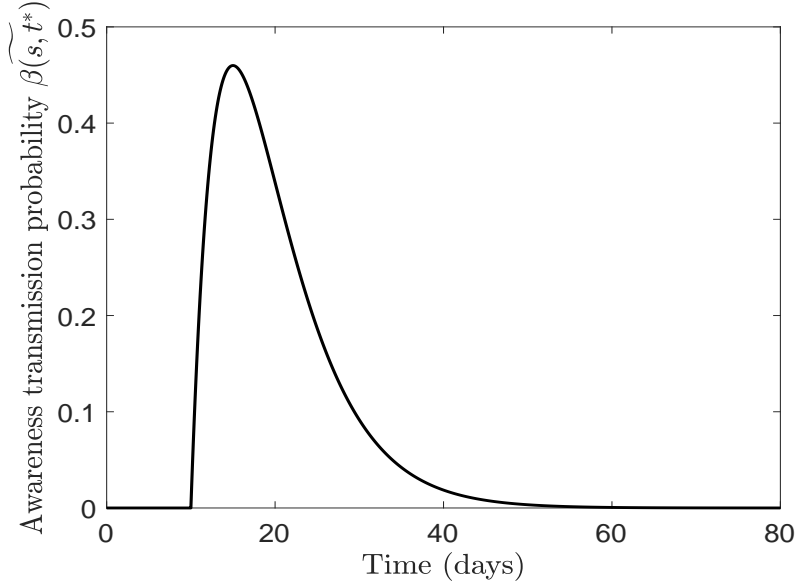}
    \caption{\textbf{Plot of the awareness probability function \pmb{$\widetilde{\beta(s,t^*)}$} for \pmb{$t^*=10$}:} Before day $t^*=10$ there is no awareness spread, $\widetilde{\beta}=0$. At day $t^*=10$ initially aware individual start informing its neighbors about infection. Through time the desire  to spread information wanes  \cite{funk2009spread}.}
    \label{beta_hat}
\end{figure}

$\widetilde{\lambda}_{jk}(s)$ denotes the probability that an aware person $S_\textbf{k}^a(s)$ will inform  an unaware neighbor  $S_\textbf{j}^u(s)$ at time $s$ via their contact: $\widetilde{\lambda}_{jk}(s)=c_{jk}\widetilde{\beta}$. The value $c_{jk}$ denotes the probability of contact between  neighbors \textbf{j} and \textbf{k}, with $\widetilde{\beta}$ denoting  the average 
probability that risk information will pass from an aware individual to an unaware neighbor. $\widetilde{\beta}$ is the average of a first increasing and then decaying and waning function  $\widetilde{\beta(s,s^*)}$ over the awareness period, that is,  $\widetilde{\beta}=\widetilde{\gamma}\int_{s^*}^{s*+\frac{1}{\widetilde{\gamma}}}\widetilde{\beta(s,s^*)}ds$, where $s^*\geq t^*$ is the first day that a person becomes aware. $\widetilde{\beta(s,s^*)}$ is a function of $s^*$, because before this time the person was unaware and therefore, incapable of spreading risk information.
Figure (\ref{beta_hat}) helps visualize  the probability function $\widetilde{\beta(s,t^*)}$ for $t^*=10$, a function acting on initially aware individual, that is,  individuals who became aware on the first day of awareness spread $t^*$.  After $10$ days, information enters the system, people learn   that an epidemic is taking place and  the risk and severity of infection. The higher the value of $P^*$ the less effective the campaign in reducing the impact of information on the outbreak.

\begin{figure}[htp]
    \centering
    \includegraphics[width=.5\textwidth]{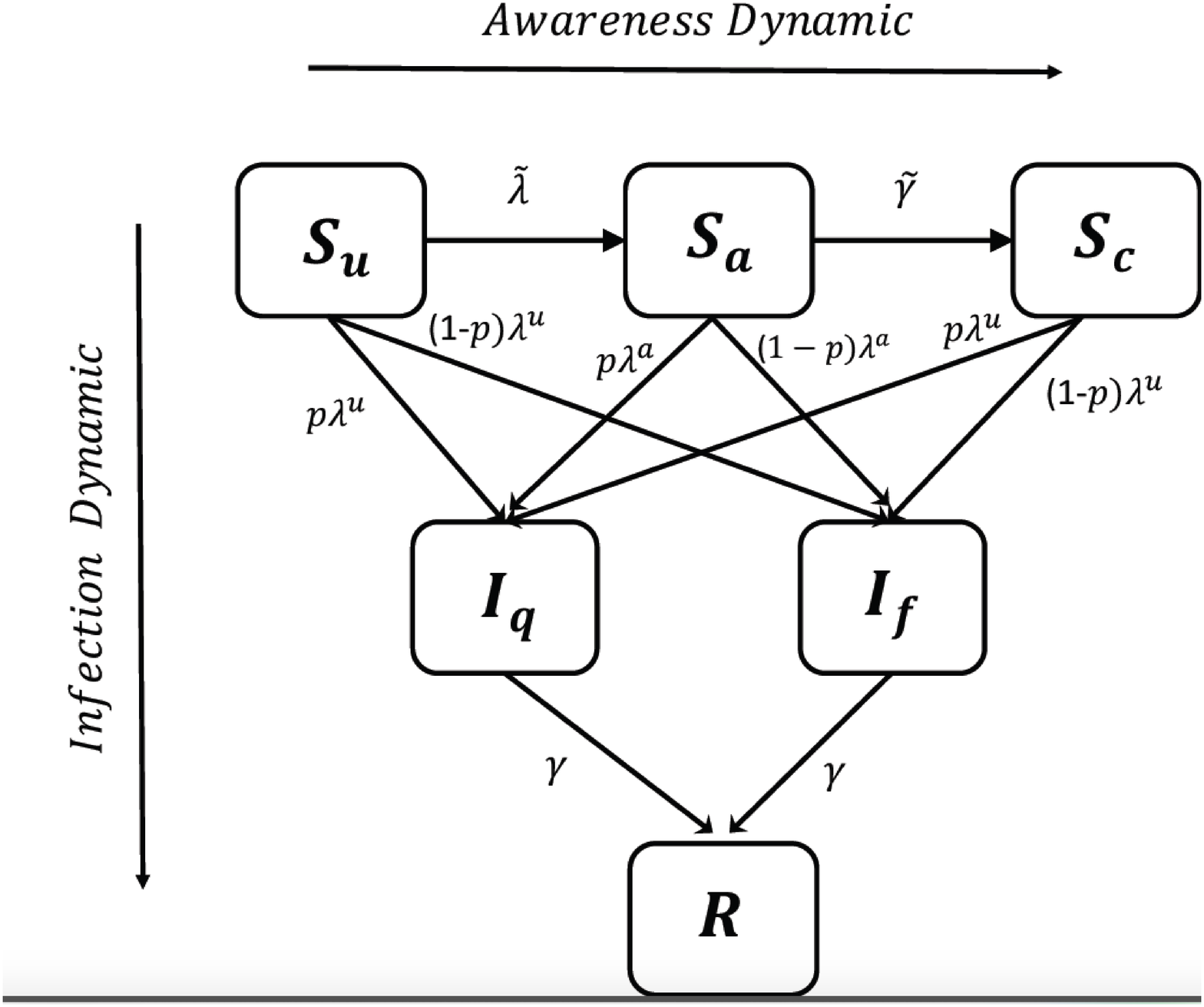}
    \caption{\textbf{Flow diagram of the analogous mean-field model:} The parameters $\widetilde{\lambda}$, $\lambda^u$ and $\lambda^a$ denote for the average force of awareness, average  force of infection for unaware, indifferent and  aware individuals respectively. The parameters $\gamma$ and $\widetilde{\gamma}$ denote  the average recovery rates from, infection and awareness, respectively. Further,  $p$  denotes the fraction of infected population that follow self-quarantine ($\pmb{I_q}$).}
    \label{fig:flow}
\end{figure}


\section{Simulations}
In this Section, we  perform our analysis and simulations on three different networks: Erd\H{o}s-R\'{e}nyi  random network $\textbf{G}_E$, Small-world network $\textbf{G}_W$, and Scale-free network $\textbf{G}_S$. These networks are defined in the appendix.
To generate the networks, we use Networkx\footnote{We use the NetworkX124 \url{https://networkx.github.io/} open software platform to generate and analyze the
network.}  and \ErdosRenyi -graph algorithm to generate  $\textbf{G}_E$ with probability of connection $p = 0.001$. Using the same package, the  Watts-Strogatz algorithm is used to generate Small-world network $\textbf{G}_W$ with probability of rewiring $p = 0.5$.
We also use Barabasi–Albert's  preferential attachment model in Networkx to generate a Scale-free
network $\textbf{G}_S$ with degree distribution $P(k)\sim k^{-2.11}$ \cite{barabasi1999emergence}. The size of all networks is 10000 and the average
number of neighbors for all networks is equal to 10.  However, the networks differ in terms of degree
distributions and other properties due to their different structures. Simulations start by infecting   the most connected node (index case) to reach a balanced initial condition \cite{azizi2018using}.  They are came out with the model baseline parameters in Table (\ref{parameters}), unless stated
otherwise.

\begin{table}[htp]
\centering
\resizebox{\columnwidth}{!}{
\begin{tabular}{ llp{11.11cm}ll}
\toprule[1.5pt]
  & \multicolumn{4}{c}{\head{}}\\
  & \normal{\head{Parameter}} & \normal{\head{Description}}
  & \normal{\head{Unit}} & \head{Baseline}\\
  \cmidrule(lr){2-5}\cmidrule(l){3-5}
  \multirow{2}{2.5cm}{Network Parameters} &  N& Number of nodes in the Network & People & $10000$\\
   &  \textbf{G}$_E$& Erd\H{o}s-R\'{e}nyi  random network  & -- & $G(N,0.001)$\\
    &  \textbf{G}$_W$& Small-world  network  & -- & $G(N,10,0.5)$\\
    &  \textbf{G}$_S$& Scale-free  random network  & -- & $P(k)\sim k^{-2.11}$\\
 &  $\bar{C}$& Average number of contact per neighbor per unit time  & contact/time & $0.75$\\
&$X(\bar{x})$& Level of education distribution (its mean)   &--&  $\beta(5,2) (0.7)$  \\
  \cmidrule(lr){2-5}\cmidrule(lr){3-5}
  \multirow{2}{1.1cm}{Infection Parameters} & $\beta$ & Probability of infection transmission per contact  & 1/contact
  & $0.011$\\
&$1/\gamma$& Average time to recover without treatment   & days&  $10$ \\
\cmidrule(lr){2-5}\cmidrule(lr){3-5}
\multirow{4}{1.1cm}{Awareness Parameters} 
& $\widetilde{\beta}$ & Average probability of awareness transmission per contact  & 1/contact
  & $0.3$\\
&$P^*$& Prevalence threshold    & 1 &  $0.1$  \\
&$1/\tilde{\gamma}$& Average time of behavior change for susceptible individuals   & days&  $7$  \\ 
&$\kappa$&  Saturation factor in $\sigma$ function& 1&  $0.85$  \\ 
&$\theta$& Parameter in which, half maximum $\sigma$ function is obtained   & 1&  $0.5$  \\
&$x^*$& Level of education threshold  & 1&  $0.4$ \\ 
\bottomrule[1.5pt]
\end{tabular}}
\caption{Parameters and their baseline values  assumed for simulations, unless stated otherwise.}
\label{parameters}
\end{table}

\subsection{Time series of infection}  

\begin{figure}[htp]
    \centering
    \includegraphics[width=.65\textwidth,height=.35\textheight]{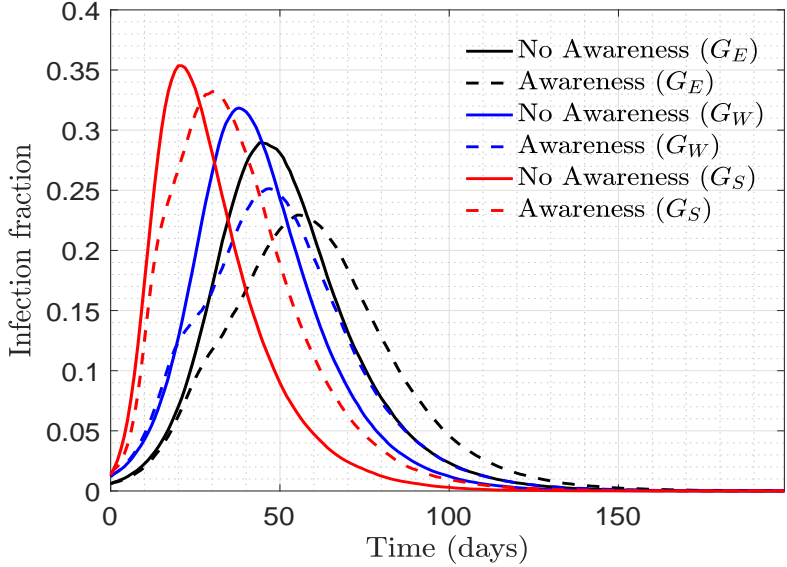}
    \caption{\textbf{Prevalence of infection versus time for three different network structure:} the curves are the mean of $100$ different stochastic simulations seeding the same initial condition. The diffusion of infection happens faster and more intense in  more heterogeneous network \textbf{G}$_S$, and  awareness diffusion has less  impact on reducing peak of infection for \textbf{G}$_S$  (by $6\%$). For other  networks reduction of peak is by $19\%$.}
    \label{scenarios}
\end{figure}

To determine the effectiveness of awareness spread, we compare the prevalence over time in the absence and presence of awareness. 
 Figure (\ref{scenarios}) shows the results for  network structures $\textbf{G}_E$, $\textbf{G}_W$, and $\textbf{G}_S$. 
We observe  that  all  networks,   with or without awareness,  support an outbreak for the chosen parameters (that is, the probability of extinction is low or the results are conditioned on non-extinction). Due to a lack of epidemic threshold for Scale-free network \cite{moreno2002epidemic, chowell2003worst,may2001infection}, we always observe an outbreak with  severity that depends on the initial infected index. For the other network structures, we observe an  epidemic threshold depends on the network topology; specifically, on the average number of neighbors and the levels of heterogeneity in the number of neighbors (mean and variance of degree distribution) \cite{kiss2017mathematics}. For example,  for the case of Small-world networks Moore et al. \cite{moore2000epidemics} derived an  analytic expression for the percolation threshold $p_c$, above which  there will be an outbreak. For the case when the network is homogeneous (\ErdosRenyi network), the epidemic threshold is proportional to the average number of neighbors (average degree) \cite{pastor2015epidemic}.  Hence,  for our simulation for both networks $\textbf{G}_E$ and  $\textbf{G}_W$ we approximated the basic reproduction number using epidemic take-offs, when  we had 
 an outbreak, that is,  whenever $\mathcal{R}_0>1$ ( probability of extinction for the parameters used seemed to be negligible).
 
The spread of infection on Scale-free network $\textbf{G}_S$ is faster than for  the other two network topologies.  Moreno et al. \cite{moreno2002epidemic} ,  modeled infection with immunity  on Scale-free and Small-world networks, they observed  that the spread of infection on Scale-free complex networks is faster  than that of Small-world networks due to a  lack of epidemic threshold for network $\textbf{G}_S$;  large connectivity fluctuations (heterogeneity in degree) on  this network  causes stronger  outbreak incidence \cite{moreno2002epidemic}. Our result is also related to  \cite{chowell2003worst}, in which Chowell et al.  tested the severity of  an outbreak using an  SIR model over  a family of Small-world networks and an Scale-free network found out that the worst case scenarios (highest infection rate) are observed in the most heterogeneous network, namely, Scale-free networks.

The speed for infection has an impact on efficiency of  awareness spread: for Scale-free network $\textbf{G}_S$ awareness reduces the peak of infection by roughly $6\%$, but  for $\textbf{G}_E$ and $\textbf{G}_W$ this reduction  is around $19\%$. Since in  $\textbf{G}_S$ network infection spreads faster,  hubs (individuals with many neighbors) get infected  faster and loosing the chance of  transmitting  or receiving  awareness. Wu et al. \cite{wu2012impact} showed a similar result: global awareness (behavior change because of higher prevalence  in the population) on Scale-free networks cannot be as effective as local awareness  in reducing infection. 

\subsection{Initiation of awareness}

 The presence and spread of awareness are coupled to the presence of infection, with awareness spreading, by design,  after the infection reaches the level $P^*$. Therefore, there is a time-lag between starting time of infection  and awareness spread. What should the value of $P^*$  be, if the goal is to reduce the final epidemic size? When the infection disperses  fast- high basic reproduction number- people must  be informed immediately, or the policy will have no effect.  As noted, there is no threshold condition for $\textbf{G}_S$ network. We did not derive  expressions for epidemic thresholds. Instead,  we used epidemic take-offs to estimate the basic reproduction number as  $\mathcal{R}_0=1.8$ and $2.01$, numbers  corresponding to $\textbf{G}_E$ and $\textbf{G}_S$, respectively.  Numbers that are  not too high- such as that  of  the $\mathcal{R}_0$ for measles \cite{keeling2011modeling}-  we are giving  awareness a chance to be effective, since an  epidemic with low $\mathcal{R}_0$ will reach the value $P^*$ much slower.

\begin{figure}[htp]
        \centering
        \begin{subfigure}[b]{0.47\textwidth}
            \centering
            \includegraphics[width=\textwidth]{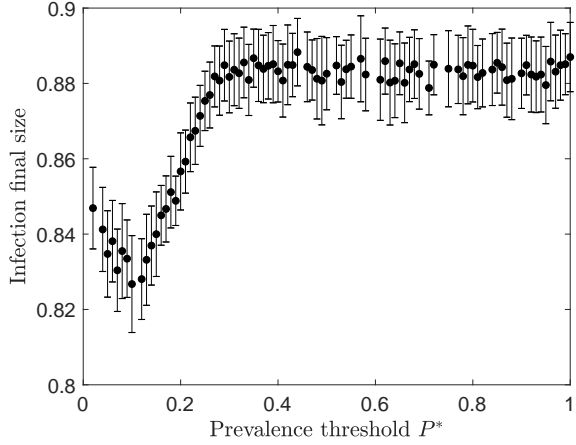}
            \caption[Network2]%
            {{\small Infection final size versus prevalence threshold for \ErdosRenyi random network \textbf{G}$_E$}.}    
            \label{IE}
        \end{subfigure}
        \begin{subfigure}[b]{0.47\textwidth}
            \centering
            \includegraphics[width=\textwidth]{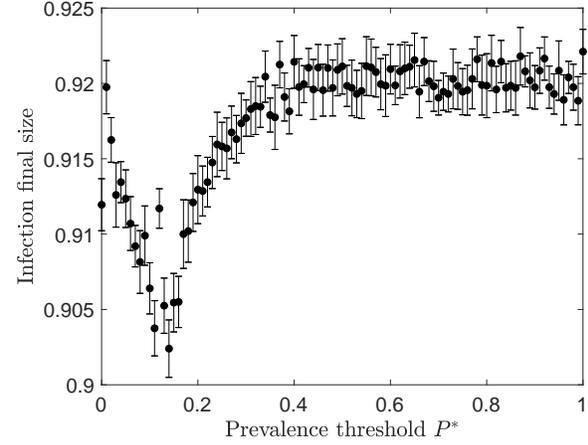}
            \caption[Network2]%
            {{\small Infection final size versus prevalence threshold for Small-world random network \textbf{G}$_W$}.}    
            \label{IW}
        \end{subfigure}
        \begin{subfigure}[b]{0.475\textwidth}   
            \centering 
            \includegraphics[width=\textwidth]{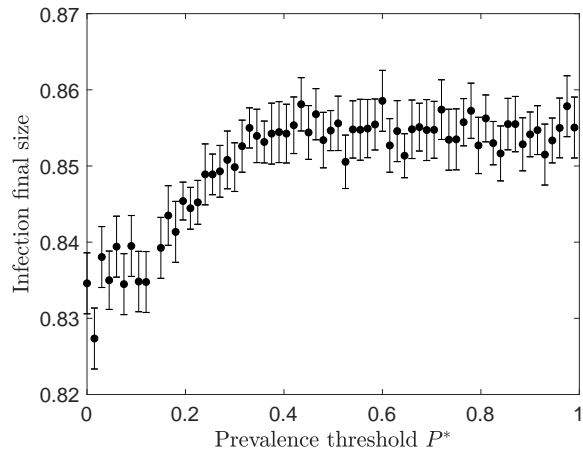}
            \caption[]%
            {{\small Infection final size versus prevalence threshold for Scale-free random network \textbf{G}$_S$}.}    
            \label{IS}
        \end{subfigure}
     \caption{\small{\textbf{Infection Final Size Versus Prevalence Threshold:} the circles are the mean of $100$ stochastic simulations and error bars are $95\%$ confidence interval. The impact of prevalence threshold  on infection final size  depends on network topology. For networks $\textbf{G}_E$ and $\textbf{G}_W$ there is an optimal $P^*$ to minimize infection final size, subfigures (\ref{IE},\ref{IW}). For the heterogeneous network $\textbf{G}_S$ the optimal point for $P^*$  disappears, subfigures (\ref{IS}).} } 
        \label{finalsize_pr}
    \end{figure}

Proceed to  quantify the sensitivity of the final size of the outbreak as a function of the prevalence threshold $P^*$ over the network structures \textbf{G}$_E$,  \textbf{G}$_W$, and \textbf{G}$_S$,  Figure (\ref{finalsize_pr}), for the simulated  \textbf{G}$_E$ network that representing  homogeneous random mixing  and  for the simulated \textbf{G}$_W$ network, we identify an optimal point for $P^*$ (the point that final size is minimized), that means awareness spread   is most effective in reducing final size  for these values of $P^*$.  For  larger values of $P^*$,  the awareness policy is useless while   for  smaller values of $P^*$ is less effective, see subfigures (\ref{IE}), and  (\ref{IW}).

For the simulated networks $\textbf{G}_S$ - in which network structure follows a power-law distribution -  we observed that the sooner  awareness gets started, the smaller the total number of infected individuals will be, see subfigures (\ref{IS}) (see also \cite{moreno2002epidemic, chowell2003worst}). 
    
\begin{figure*}[htp]
    \centering
          \includegraphics[width=.65\textwidth,height=.35\textheight]{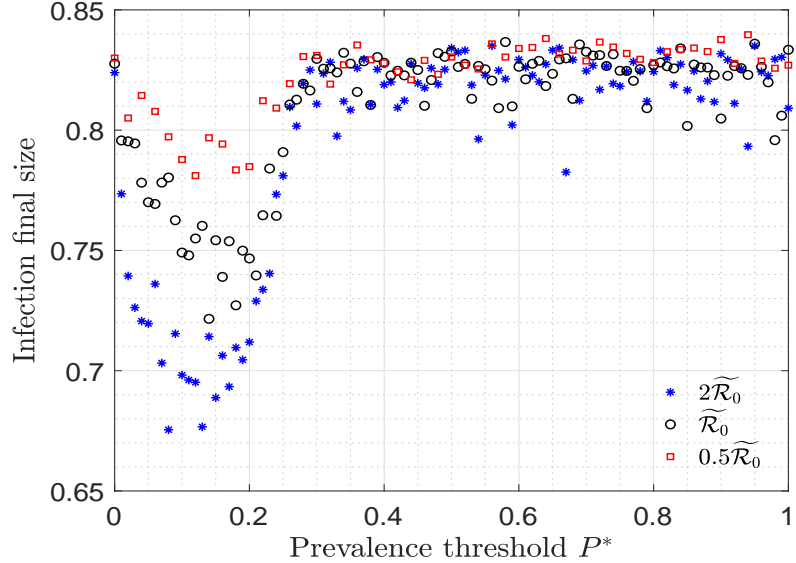}
 \caption{\textbf{Infection final size versus \pmb{$P^*$} for different awareness \pmb{$\widetilde{\mathcal{R}_0}$}}: Increasing $\widetilde{\mathcal{R}_0}$ via increasing the average period   of awareness will make the impact of prevalence threshold on final size stronger.}
  \label{rhat}
\end{figure*}

  To investigate the impact of awareness on the previous result and on network $\textbf{G}_E$, we vary the value of the awareness basic reproduction number $\widetilde{\mathcal{R}_0}$ via changes in the average time of behavior change $1/\tilde{\gamma}$.  Figure (\ref{rhat}) shows that  the value of the optimal prevalence threshold $P^*$ increases as the  $\widetilde{\mathcal{R}_0}$ increases.  Accelerating awareness spread- increasing $\widetilde{\mathcal{R}_0}$- does not seem to change final awareness size  trend (Figure (\ref{finalsize_pr})).  Nevertheless, it makes the prevalence threshold $P^*$ more effective  at  reducing infection final size.
  
  \section{Discussion} 

We used an agent-based network model to model the spread of infection and risk information within a population of individuals in different epidemiological states, education levels and information states. In our model infection spreads within an SIR framework while awareness disseminates only among susceptible individuals. Specifically, when a fraction of the infected people reaches  a pre-determined threshold, public health officials  start a massive information campaign on the risk and severity of infection.  Information  is transmitted by those not infected only. Awareness dynamics are triggered by public health authorities with the objective of conveying information about risk and severity of infection via aware susceptible individuals to  unaware susceptible neighbors in the network. The level of awareness achieved is tied in to the the educational level of each individual and the disease prevalence. Aware susceptible individuals  can become indifferent, stop propagating information on the the risk of infection, after sometime.

Simulations of the spread of infection or awareness are carried out on different  network topologies. The patterns vary even when preserving some  network properties such as the mean degree \cite{chowell2003worst, shirley2005impacts, moreno2002epidemic}. We simulated the  spread of awareness and disease over three different network structures, namely \ErdosRenyi ,  Small world  and  Scale-free networks. 

The results of the infection-awareness model show that Scale-free networks support a fastest rate of infection with the impact of awareness having the lowest effect on transmission. Small-world  and  \ErdosRenyi networks  became less effective at disease transmission, and  the role of information via aware individuals had a stronger impact in reducing transmission, Figure (\ref{scenarios}). These result are somewhat similar  to those obtained by Chowell et al. \cite{chowell2003worst} and  Moreno et al. \cite{moreno2002epidemic} as they studied the rate of infection rate of growth on an SIR model over  Small-world  and  Scale-free networks, in the absence of awareness. These researchers  observed  that the spread of infection on Scale-free complex networks is faster  than in a Small-world one. Pastor et al. \cite{pastor2002immunization} and Dezs{\H{o}} et al. \cite{dezsHo2002halting}  focused on identifying the best  strategy that can be used to control (reduce) an epidemic peak on a scale-free network. They concluded that to in order to control infection public health workers needed to immunize the hubs at an early stage. Our results are somewhat related, that is, we find that in order to control an  infection, public health workers need to start awareness spread  soon, that is,  before  hubs play the role of   super-spreaders. 
    
Infection final size is a function of network topology and prevalence threshold, $P^*$, and the basic reproduction number $\mathcal{R}_0$. We observed that  for \ErdosRenyi and Small-world networks under  small value of $\mathcal{R}_0$, an optimal $P^*$ can be found for which the infection final size is minimized. This optimal point is the value that provides the best outcome  under the prevalence $P^*$ at the  time, $t^*$,  the time when  temporary protection was promoted by aware susceptible individuals, Figures (\ref{IE}) and (\ref{IW}). For Scale-free network, for which there is no epidemic threshold,  a monotonic increasing function final size is observed,  a function of the prevalence threshold $P^*$,  there is no  optimal point $P^*$ that minimizes infection final size, Figure (\ref{IS}).

    \underline{\textbf{Infection network and awareness network:}} One of the big assumptions in our model is that the infection and awareness spread over the same network,  ignoring the fact that individuals may have multiple sources of information outside their physical contact network. In fact, we know that information does flow through  virtual neighbors such as facebook friends. How should we incorporate the role of  physical network  and virtual information? We are  exploring possibilities.

    \underline{\textbf{Static network:}} Again, since we are focusing on single epidemic outbreak, we have assumed a  short time frame in our simulation. We also assumed that  the context network is static.  The need to extend the model to a dynamic network  where individuals are allowed to change neighbors is important in the study of disease spread over the longer time scale. The impact of awareness in such extended model would be prone to change as well. \\

For sexually transmitted infections on Scale-free  network \cite{liljeros2001web}, waiting for enough people to become  infected people before starting a campaign about risk of infection is not reasonable. For measles in children with   basic reproduction numbers  very high \cite{keeling2011modeling}, waiting to reach pre-selected prevalence thresholds may also not be great idea. Our model results,  highlight the importance of  campaigns that warns a population about the risk and severity of infection for diseases that do not spread too fast, possibly including some flu  infections  or the severe acute respiratory syndrome (SARS). For disease with a basic reproduction number that it is is not high \cite{keeling2011modeling}, and assuming that these infections spread over a random homogeneous networks such as \ErdosRenyi , it may not be unreasonable   the  existence of  an optimal prevalence threshold to start and starting  an awareness dynamic campaign.


\clearpage
\newpage
\section*{Appendix}
In this appendix, first we define some models for the network structures we used for our simulation, then we present the pseudocode of our model. 
\begin{definition}{\ErdosRenyi random network}
Suppose we have   $n$ disconnected nodes, and then we  make edge between   two arbitrary nodes  with probability $p$ independent from every other edge. The graph constructed via this model is called \ErdosRenyi random network $G(n,p)$. The parameter $p$ in the model is a  weighting function $p\in\left[0,1\right]$, where for $p$ closer to one  the generated graph is  more likely to construct graphs with more edges \cite{erdds1959random}. 
\end{definition}

\begin{definition}{Small-world  network}
Suppose we have $n$ nodes over a ring, where  each node in the ring is connected to  its $k$ nearest neighbors in the ring. Then with probability $p$ and independent from any other pair of edges we select two edges ${i_1j_1}$ and  ${i_2j_2}$ such that $j_k$ is from the k nearest neighbor of $i_k$ for $k=1,2$, and  rewire them. Then the constructed network is called Watts-Strogatz Smal-world network \cite{watts1998collective} $G(n,k,p)$.
\end{definition}

\begin{definition}{Preferential attachment Scale-free  network}
Suppose we want to generate a network of $n$ nodes. We start with one node and follow the following procedure $n$ times:  the probability of attaching a new node to  the existing ones is proportional to their current degree: \\$P(\textnormal{connect a new node i to existing node j})= \frac{deg(j)}{\sum_k(deg(k))}$, where the $deg(k)$ is the number of neighbors for node $k$, then with this probability we make new edge $ij$. The constructed network via this model is called  Barabasi–Albert Scale-free network \cite{barabasi1999emergence}. The degree distribution of this network follows a power-law distribution.
\end{definition}

Now, we define assumptions and algorithm of our model. Person $i$ at any given day $t$ has one of the states: $S^u, S^a, S^i, I^q, I^f$, or $R$. The fraction of infected people at dat $t$ is $P(t)$, and $P^*$ is a positive fraction $\in [0,1]$. We define $t^*=min_t\{ P(t)=P^*\}$ which is the first time that $P(t)$ reaches $P^*$. We assume Education level for each person $k$, $x_k$ does not change by time. Also prior to time $ t^*$ every susceptible person is at state $S^u$, and finally the time scale for each update is day, in which each person can have at most one update. The following algorithms show the details. 
\begin{table}[hpt]
\centering
\begin{tabular}{lp{10cm}}
\toprule[1.5pt]
   \textbf{Notation} & \textbf{Description}\\
  \cmidrule(lr){1-2}
  $S_t$ &  Set of susceptible  nodes  at time t\\
    $U_t$ &  Set of unaware  nodes  at time t\\
      $A_t$ &  Set of aware  nodes  at time t\\
      $C_t$ &  Set of careless nodes  at time t\\
   $I_t$ &  Set of infected  nodes  at time t\\
   $Q_t$ &  Set of quarantine  nodes  at time t\\
   $F_t$ &  Set of free  nodes  at time t\\
    $R_t$ &  Set of recovered nodes  at time t\\
    $G.N(\textbf{k})$ &  Set of neighbors of node \textbf{k} in Network G\\
  $IP$(\textbf{k})& Infection period for infected node \textbf{k}\\
  $AP$(\textbf{k})& Awareness period for aware node \textbf{k}\\
    $ern(\alpha)$ & Exponential random number with average $\alpha$ for $\alpha>0$\\
    $urn$ & Uniform random number in $[0,1]$\\
    $c_{kj}$ & Probability of having contact between two neighbors \textbf{k} and \textbf{j}\\
    $A\xrightarrow[~]{\textbf{k}}B$& Element \textbf{k} moves from set A to set B\\
\bottomrule[1.5pt]
\end{tabular}
\caption{Table of notation for a conventional network $G$ in algorithms.}
\label{notation}
\end{table}	

\clearpage
\newpage
\begin{minipage}{1.0\linewidth}
\begin{algorithm}[H]
 \SetAlgoLined
$\hat{\sigma_k}=1_{\Big|\textbf{k}\in Q_t}$\; 
$\sigma_j= \frac{ \kappa x_j}{(\theta+x_j)(1+pr(t))}_{\Big|\textbf{j}\in A_t}+1_{\Big|\textbf{j}\notin A_t}$\;
               
              $\Sigma_{kj}\defeq \hat{\sigma_k}\sigma_j$
 \caption{ The probability of contact per day between two neighbors: infected \textbf{k} and  susceptible \textbf{j}.}\label{help}
\end{algorithm}
\end{minipage}

\begin{minipage}{1.0\linewidth}
\begin{algorithm}[H]
 \SetAlgoLined
 $\%==============${\textit{/*Infection dynamic*/}}$==============\%$\;
\For{\textbf{k}$\in I_t$}
{\uIf{$IP(\textbf{k})=-1$}
           {$IP(\textbf{k})\leftarrow max\{1,ern(\gamma$)\}\; }
              \uElseIf{$IP(\textbf{k})>0$}
              {$IP(\textbf{k})\leftarrow IP(\textbf{k})-1$\;}
              \Else{$I_t\xrightarrow[]{\textbf{k}}R_{t+1}$\;}
            \For{$j\in G.N(\textbf{k}) \cap S_t$}  
            { \If{urn$\leq \beta\sigma_{kj}(t)c_{kj}$}
       {{$S_t\xrightarrow[~]{\textbf{j}}I_{t+1}$,~~$IP(\textbf{\textbf{j}})\leftarrow -1$ \;}
      \lIf{$x_\textbf{j}>x^*$}
      {$\textbf{j}\rightarrow Q_{t+1}$}}
    }
           }
$\%==============${\textit{/*Awareness dynamic*/}}$==============\%$\;
\For{\textbf{k}$\in A_t$}
{\uIf{$AP(\textbf{k})=-1$}
           {IP(\textbf{k})$\leftarrow$ max\{1,ern($\widetilde{\gamma}$)\}\; }
              \uElseIf{$AP(\textbf{k})>0$}
              {$AP(\textbf{k})\leftarrow AP(\textbf{k})-1$\;}
              \Else{$A_t\xrightarrow[~]{\textbf{k}}C_{t+1}$\;}
            \For{$j\in G.N(\textbf{k}) \cap U_t$}  { \lIf{ \textnormal{ j is not infected next time $t+1$ and } urn$\leq \widetilde{\beta}c_{kj}$}
       {$U_t\xrightarrow[~]{\textbf{j}}A_{t+1}$ ~~$AP(\textbf{j})\leftarrow -1$}
}
}
 \caption{ Dynamics within a typical day t}\label{main}
\end{algorithm}
\end{minipage}

\end{document}